\documentclass[preprint,12pt]{elsarticle}
\usepackage{amssymb}
\usepackage{graphicx}
\journal{Physica A}

\begin{document}

\begin{frontmatter}

\title{Leverage Bubble}
\author[ETH]{Wanfeng Yan}
\ead{wyan@ethz.ch}
\author[ETH]{Ryan Woodard}
\ead{rwoodard@ethz.ch}
\author[ETH,SFI]{Didier Sornette\corref{cor}}
\cortext[cor]{Corresponding author. Address: KPL F 38.2, Kreuzplatz
5, ETH Zurich, CH-8032 Zurich, Switzerland. Phone: +41 44 632 89 17,
Fax: +41 44 632 19 14.}
\ead{dsornette@ethz.ch}%
\ead[url]{http://www.er.ethz.ch/}%

\address[ETH]{Chair of Entrepreneurial Risks, Department of
  Management, Technology and Economics, ETH Zurich, CH-8001 Zurich,
  Switzerland}%
\address[SFI]{Swiss Finance Institute, c/o University of Geneva, 40
  blvd. Du Pont dArve, CH 1211 Geneva 4, Switzerland}%

\begin{abstract}
  Leverage is strongly related to liquidity in a market and lack of liquidity
  is considered a cause and/or consequence of the recent financial crisis.  A
  repurchase agreement is a financial instrument where a security is sold
  simultaneously with an agreement to buy it back at a later date.  Repurchase
  agreements (repos) market size is a very important element in calculating the
  overall leverage in a financial market.  Therefore, studying the behavior of
  repos market size can help to understand a process that can contribute to the
  birth of a financial crisis.  We hypothesize that herding behavior among
  large investors led to massive over-leveraging through the use of repos,
  resulting in a bubble (built up over the previous years) and subsequent crash
  in this market in early 2008.  We use the Johansen-Ledoit-Sornette (JLS) model of
  rational expectation bubbles and behavioral finance to study the dynamics of
  the repo market that led to the crash. The JLS model qualifies a bubble by the
  presence of characteristic patterns in the price dynamics, called
   log-periodic power law (LPPL) behavior. We show that there was significant
  LPPL behavior in the market before that crash and
  that the predicted range of times predicted by the model for the end of the
  bubble is consistent with the observations.
\end{abstract}


\end{frontmatter}

\section{Introduction}

Financial bubbles play a huge role in the global economy, affecting
hundreds of millions of people yet, until recently, the existence of
such bubbles, much less their effects, have been ignored at the
policy level.  Finally, only after this most recent historical
global financial crisis (which is still ongoing), officials at the
highest level of government and academic finance have acknowledged
the existence and importance of identifying and understanding
bubbles.  No less than the President of the Federal Reserve Bank of
New York, William C. Dudley, stated just a few months ago
\cite{dudley20100409}:
\begin{quotation}
  $\ldots$what I am proposing is that we try---try to identify bubbles in real
  time, try to develop tools to address those bubbles, try to use those tools
  when appropriate to limit the size of those bubbles and, therefore, try to
  limit the damage when those bubbles burst.
\end{quotation}
Such a statement from the New York Fed---representing, essentially,
the monetary policy of the United States governmental banking
system---would have been, and, in some circles, still is, unheard
of.  This, in short, is a bombshell and a wake-up call to academics
and practitioners.  Dudley proposes to ``try to develop tools to
address...bubbles''.  Before discussing tools that could address
bubbles, we must know what a bubble is.

A financial bubble is a curious beast: its meaning is accepted as
beyond obvious by average people yet its very existence is loudly
debated in angry terms among experts.  Arguably, almost any given
adult met on the street in, say Asia, Europe or North America (and,
or course, elsewhere), would know exactly what one is and could cite
examples in recent and distant history.  The dot.com bubble ending
in 2000 and the housing bubble recently ended would most likely be
the most common examples given.  More well-read---but still
non-expert---people could cite the Dutch tulip mania in the 1600s
and the South Sea Company of the 1700s.  After that, the examples
are less well-known but not because of their scarcity but just
because most people are not interested in financial history and
debates.  This is changing.

While the general population accepts bubbles, academics and
policy-makers have a decades-old tradition of arguing about whether
bubbles even exist.  This single fact could very well be the most
striking and unbelievable statement ever transmitted from the Ivory
Tower.  In spite of the lack of academic consensus on a definition,
we can give a qualitative ideas of bubbles and apply our
quantitative approach in the next sections.  Following
\cite{CaseShiller}, the term ``bubble'' refers to a situation in
which excessive public expectations of future price increases cause
prices to be temporarily elevated. For instance, during a housing
price bubble, homebuyers think that a home that they would normally
consider too expensive for them is now an acceptable purchase
because they will be compensated by significant further price
increases. They will not need to save as much as they otherwise
might, because they expect the increased value of their home to do
the saving for them. First-time homebuyers may also worry during a
housing bubble that if they do not buy now, they will not be able to
afford a home later. Furthermore, the expectation of large price
increases may have a strong impact on demand if people think that
home prices are very unlikely to fall, and certainly not likely to
fall for long, so that there is little perceived risk associated
with an investment in a home.

In this paper, instead of a housing bubble, we argue that there was
a leverage bubble that peaked and crashed in early 2008 after
building up for the years beforehand.  As we explain below, the
leverage bubble formed and grew for the same reasons as described in
the housing bubble example above: investors were afraid that if they
did not extend their leverage (buy a house) then they would lose
money later.  Further, we argue that the size of the market in
\emph{repurchase agreements} (or repos, for short) is an observable
proxy of leverage in the financial system.  We will elaborate on
repos below, but, briefly, a repo is simply a cash transaction for
an asset combined with a forward contract to buy the asset back at a
later time (hence `re-purchase'). By measuring the size of the repos
market and applying an appropriate bubble model, we can see that the
leverage crash in early 2008 was potentially a forecastable event.

The paper is constructed as follows: in Section 2, we discuss the
relationship between repos market size and the overall leverage of
the market. In Section 3, we briefly introduce the
Johansen-Ledoit-Sornette model \cite{jsl,Johansen2000f} of bubbles
and apply it to total repos market size to make an ex-post forecast
of the crash in early 2008. We conclude in Section 4.

\section{Repos market size represents the leverage of the market}

A repurchase agreement (repo) is the sale of securities together
with an agreement for the seller to buy back the securities at a
later date \cite{repowiki}. In other words, it is a contract
obliging the seller of an asset to buy back the asset at a specified
price on a given date. Therefore, a repo is equivalent to a cash
transaction combined with a forward contract.  The cash transaction
results in transfer of money to the borrower in exchange for legal
transfer of the security to the lender, while the forward contract
ensures repayment of the loan to the lender and return of the
collateral of the borrower.

To understand the possible role of repos in the generation of a
bubble, we first discuss the relationship between leverage and
balance sheet size.  We start with a very simple case, taken from
Section 2 of \cite{liquidityleverage}. Assume that an investment
bank has 100 USD in securities while its shareholder equity is 20
USD and its debt is 80 USD. Then the balance sheet of this bank
looks like:
\begin{center}
  \begin{tabular}{@{}ll}
    \hline
    Assets&Liabilities\\\hline
    Securities, 100& Equity, 20\\
    &Debt, 80\\\hline
  \end{tabular}
\end{center}
Now the leverage of the bank is:
\begin{equation}
  \frac{assets}{equity} = \frac{100}{20} = 5.
\end{equation}
Suppose that the debts of this bank are all long term debts and,
therefore, we can assume that the debt remains the same in the
balance sheet over the short period of time considered in the
argument.  Now assume that the prices of the securities increase by
10\%, so that the new balance sheet is:
\begin{center}
\begin{tabular}{@{}ll}
\hline Assets&Liabilities\\\hline
Securities, 110& Equity, 30\\
 &Debt, 80\\\hline
\end{tabular}
\end{center}
The leverage, then, becomes:
\begin{equation}
  \frac{assets}{equity} = \frac{110}{30} = 3.67 < 5.
\end{equation}
This shows that the leverage decreases as the assets' prices
increase.

However, to an investor during the bull market, reduction of the
leverage means losing money. Consider another example to demonstrate
this.  Suppose that two people $A$ and $B$ both have a house worth
1000 USD.  Assume that they somehow know that the price of gold, for
instance, will definitely increase in the near future.  Each of them
can use her house as collateral and get a maximum 2000 USD load from
a bank (based on the recent convention of poor underwriting
requirements).  Investor $A$, being somewhat unsure of her future
ability to repay her debts, applied for and received `only' 1500
USD, which corresponds to a leverage of 1.5. Investor $B$, though,
with no such qualms, asked for and received the maximum value of
2000 USD, for a leverage of 2.  Both investors used all of the
borrowed money to buy gold. After one month, the gold price, as
expected, increased by 20\%.  Both $A$ and $B$ sold all of the
leveraged gold and repurchased their respective houses for 1000 USD
(ignore interest rate for simplicity). Investor $A$ has made a
profit of 800 USD but investor $B$, the bold risk-taker, has made
almost double the profit of 1400 USD by simply increasing her
leverage by one-third.  In a sense, investor $A$'s weak-kneed
approach lost 600 USD due to failure to maximally leverage her
position.

With this lesson in mind, let us now return to the investment bank.
During the bull market, banks believe that the markets will continue
to increase and that all of their competitors will be maximally
leveraged to take advantage of the expected rise. If a bank
decreases its leverage, it means it will lose money in the future
so, guided by the practice of maximizing short-term profits by any
means necessary, banks increase their leverage in order to get more
return in the future. How large they will increase their leverage
depends on their expectation of the future market. If the market
performs very well now, they expect that the future will be very
good, also. This means that they will change their leverage based on
the return \emph{now}.  Regardless of whether this is a good thing
or not, for our study, we can use this because it implies that the
total asset growth should be proportional to the leverage growth.
This is demonstrated in Fig. 8 of \cite{liquidityleverage}. In that
paper, the authors used quarterly data from more than 10 years for
six major U.S. investment banks: Lehman Brothers, Merrill Lynch,
Morgan Stanley, Bear Stearns, Goldman Sachs and Citigroup Markets.
The total asset growth of the banks is found strongly proportional
to the leverage growth.  So we know that when the expectation of the
market is high, the investment banks tend to increase their
leverage. The next question, then, is: how can a bank change its
leverage?

Repos play a key role here. A typical balance sheet of an investment
bank has not only the long term debt but also repos. Therefore, a
typical balance sheet is as follows:
\begin{center}
\begin{tabular}{@{}ll}
\hline Assets&Liabilities\\\hline
Trading assets&Repos\\
Reverse repos&Long term debt\\
Other assets&Equity \\\hline
\end{tabular}
\end{center}
Recall that a repo is the sale of securities together with an
agreement for the seller to buy back the securities at a later date.
Long term debt is normally a small fraction of the balance sheet and
can be assumed to be constant over the time scale of interest here
(a few years at most). In this case, when banks want to increase or
decrease their leverage, they will write repos.

One may argue that the haircut of the repo\footnote{The ``haircut is
the difference between the true market value of the collateral and
that used by the dealers in the repo contract. This haircut reflects
the underlying risk of the collateral and protects the buyer against
a change in its value. Haircuts are therefore specific to classes of
collateral.} is also a very important role for the leverage of the
banks. We completely agree with this and the repurchase haircut
should be counted here. However, the historical data shows that the
haircut remains approximately within a range between 10\% and 20\%
during `normal' (i.e., non-crisis) times. During a financial crisis,
the haircut will rise sharply to a very high level. When there is a
shortage of liquidity, for instance, during the recent financial
crisis, investors are afraid to trade. Increased haircuts and
decreased repos size usually occur simultaneously. In this paper, we
want to investigate the question of whether or not the dynamics of
repos activity shows any precursory information before a large
crash.  Of course, this means we only use data before a crash to try
to estimate the time of its onset. Since the haircut is almost
constant for a long time before a crash, all of the leverage
information lies in the repo size of the market.

To summarize this section, we claim that:
\begin{enumerate}
\item investors want to increase their leverage when their expectations of
  future gains of the market increase;
\item they will use repos to increase their leverage;
\item therefore, the total repo market size is a proxy to measure the overall
  expectation of all investors.
\end{enumerate}

\section{Predicting financial crashes with the Johansen-Ledoit-Sornette model}

In the last section, we said that the repos market size represents
the average leverage of the market and the leverage represents the
investors' expectation of future market returns.  We now discuss how
the dynamics of leverage among traders could lead to a bubble and
how this bubble can be identified as it grows.

We have argued before that bubbles are the result of imitation and
herding behavior among investors
\cite{Sornette1996,jsl,Johansen2000f,sornettecrash}. In the current
case, investors increase their leverage when they see others doing
so because, as discussed above, they think that they will lose money
if they are the only ones not taking this strategy.  Of course, this
is a self-reinforcing (positive feedback) process: the numbers of
leveraged investors and their levels of leverage will increase in a
game of financial copycat.  At some point, though, some investors
are bound to notice that the numbers are too large and they will
start to deleverage.  Others nervously waiting for this signal will
unload as well and the bottom will drop out.  When this occurs, the
repo market size goes down dramatically and the haircut of the repo
increases very sharply, both leading to rapid loss of liquidity in
the repo market.

This qualitative process is quantified in the
Johansen-Ledoit-Sornette (JLS) model to describe the herding
dynamics during a bubble \cite{jsl,Johansen2000f}. This model
combines the economic theory of rational expectation bubbles,
behavioral finance on imitation and herding of investors and traders
and the mathematical and statistical physics of bifurcations and
phase transitions. Many successful predictions of financial market
crashes based on this model have been made, such as the 2006--2008
oil bubble \cite{oil}, the Chinese index bubble in 2009 \cite{ssec},
real estate market in Las Vegas \cite{vegas}, South African stock
market bubble \cite{southafrica}.  Also, new methods using this
model to predict stock market rebounds rather than the crashes are
being developed \cite{prrebound}.

In the JLS model, (the logarithm of) price is used as a proxy for
herding behavior among traders (see \cite{js} for justifications on
the use of log-price versus price).  Since we argue that the repo
market size is also a proxy for herding via the leverage level, we
substitute it for the log-price in the JLS model.  For the total
repos market size $R(t)$ at time $t$, we use the following JLS model
specification (corresponding to replace log-price by repos volume in
the JLS equation):
\begin{equation}
  R(t) = A + B|t_c - t|^m + C|t_c - t|^m \cos(\ln \omega|t_c - t| + \phi) ~,
  \label{eq:jls}
\end{equation}
where $t_c$ is the crash time and $m, \omega, \phi, A, B$ and $C$
are parameters. To determine the values of these parameters, we want
to minimize the sum of squares:
\begin{eqnarray}
&(t_c, m, \omega, \phi, A, B, C) =\\\nonumber &\arg \min \sum_t
(R(t) - A + B|t_c - t|^m + C|t_c - t|^m \cos(\ln \omega|t_c - t| +
\phi))^2~.
\end{eqnarray}
We hypothesize that the run-up to the sudden large drop in the repos
market in early 2008 was characterized by LPPL dynamics, supporting
our claim of the entanglement of expectations, leverage and herding
behavior.

To test this hypothesis, we use the weekly data of US primary
dealers' total repos size from 6 July 1994 to 23 June
2010.\footnote{We thank Tobias Adrian
  from the Federal Reserve Bank of New York for providing the data.}\footnote{The
  primary dealers list: BNP Paribas Securities Corp, Banc of America Securities
  LLC Barclays Capital Inc, Cantor Fitzgerald \& Co, Citigroup Global Markets
  Inc, Credit Suisse Securities (USA) LLC, Daiwa Capital Markets America Inc,
  Deutsche Bank Securities Inc, Goldman, Sachs \& Co, HSBC Securities (USA)
  Inc, Jefferies \& Company, Inc, J.P. Morgan Securities LLC, Mizuho Securities
  USA Inc, Morgan Stanley \& Co, Incorporated Nomura Securities International,
  Inc, RBC Capital Markets Corporation RBS Securities Inc, UBS Securities LLC.}
The data have very strong seasonal effects due to the fact that
banks try to remove their repos to improve their balance sheet at
the end of each quarter. To remove the seasonal effect, we used a 13
week (1 quarter) moving average.

We fit this smoothed time series with the JLS equation
(\ref{eq:jls}) in time windows defined by $(t_1, t_2)$.  We chose a
fixed $t_2 =$ 13 February 2008, approximately one month before the
observed peak of the repos volume. We then repeated the analysis
with an ensemble of 7 values of $t_2$, each separated by 7 days for
the 3 weeks before and after 13 February 2008.  Note that the 7
values of $t_2$ bracket a time span of 6 weeks, with the end of that
period (5 March 2008) being just before the large drop in the repos
market.  Also note that an observer in the past on this date would
not have noticed any unusual drop in the time series.  That is, the
impending crash was not obvious based on any recent trend in the
data (though perhaps some market intelligence could have provided an
indication).  For each value of $t_2$, we use an ensemble of
different $t_1$'s.  Each ensemble brackets a range between 6 and 18
months before the respective $t_2$ and values of $t_1$ are separated
by 7 days.

The fit for a particular $(t_1, t_2)$ interval is generated in two
steps. First, the linear parameters $A, B$ and $C$ are slaved to the
non-linear parameters by solving them analytically as a function of
the nonlinear parameters.  We refer to \cite{jsl} (page 238 and
following ones), which gives the detailed equations and procedure.
Then, the search space is obtained as a 4 dimensional parameter
space representing $m, \omega, \phi$ and $t_c$. A heuristic search
implementing the Taboo algorithm \cite{ck} is used to find 10
initial estimates of the parameters which are then passed to a
Levenberg-Marquardt algorithm \cite{kl,dm} to minimize the residuals
(the sum of the squares of the differences) between the model and
the data. The bounds of the search space are:
\begin{eqnarray}
m &\in& [0.001, 1.999]\\
\omega &\in& [0.01, 40]\\
\phi &\in& [0.001, 2\pi-0.001]\\
t_c &\in& [t_2, t_2 + 0.375(t_2 - t_1)]
\end{eqnarray}

\begin{figure}[htp]
\centering
\includegraphics[width=0.9\textwidth]{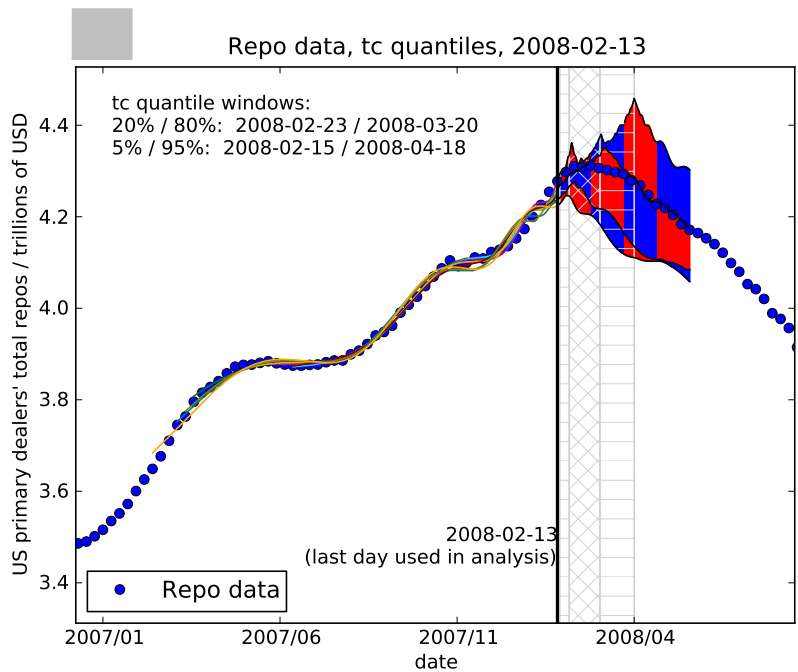}
\caption{Results of the calibration of the JLS model calibrated to
the time evolution of the total repos market size. The end time of
the time series is fixed to $t_2=$2008-02-03, shown as the dark
vertical line to the left of the quantile windows. For different
starting time, the probability density of the crash time $t_c$ is
shown in quantiles. The curves on the right of the dark vertical
line are the extrapolated quantile repos volume, which are found
consistent with the realized trajectory of the moving average of the
repos market size. } \label{fg:fixt2fitresult}
\end{figure}

Fig.\ref{fg:fixt2fitresult} shows the fitting results with a fixed
end of the time series $t_2$ = 3 February 2008 and the ensemble of
$t_1$s as described above.  The use of many fits provides an
ensemble of $t_c$'s, from which we can calculate quantiles of the
most likely date of a crash.  The 20\%-80\% quantile region is shown
on the figure as the inner vertical band with diagonal
cross-hatching.  The 5\%-95\% quantiles are shown as the outer
vertical band with horizontal hatching.  The dark vertical line to
the left of the quantile windows represent the last observation used
in the analysis, that is, $t_2$. The shaded envelopes to the right
of $t_2$ represent 20\%-80\% and 5\%-95\% quantiles of the
extrapolations of the fits.  From the plot, we see that both the
$t_c$ quantiles and the extrapolation quantiles are consistent with
the observed trajectory of the moving average of the repos market
size.

\begin{figure}[htp]
\centering
\includegraphics[width=0.9\textwidth]{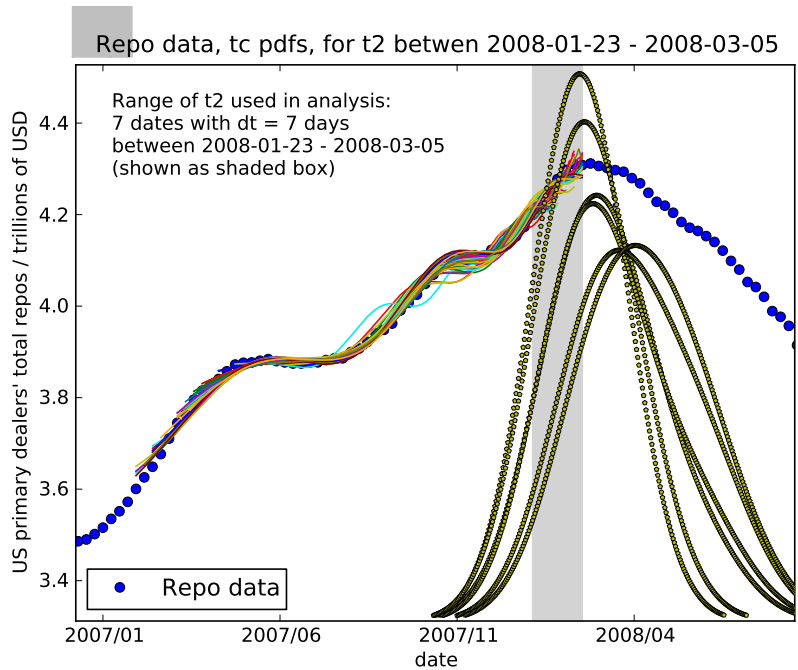}
\caption{We vary the end of the window $t_2$ within the grey area
and show the probability density of the crash time $t_c$ for each of
these $t_2$, as the bell-shaped curves with open circle symbols.}
\label{fg:t2neartcpdfs}
\end{figure}

Our use of 7 values of $t_2$'s in the 6 week window described above
is to address the issue of the stability of the predicted crash time
in relation to $t_2$.  We fit the ensemble of $(t_1, t_2)$ intervals
as described above and plot the pdf's of the predicted crash time
$t_c$ for each $t_2$. The result is shown in Fig.
\ref{fg:t2neartcpdfs}. From the plot, one can observe two regimes.
The first four pdf's corresponding to the earliest $t_2$'s peak
practically at the same value, showing a very good stability. The
last two pdf's show a tendency to shift to the future, as some of
the used data starts to be sensitive to the plateauing of the repos
volume. Overall, the observed stability of the predicted
distributions of $t_c$'s means the calibration of the JLS model is
quite insensitive with when the prediction is made. This is proposed
as an important validation step for the relevance of the JLS model.
This suggests that the JLS model can be used for advance diagnostic
of impending crashes. The present results add to those accumulating
within
 the ``financial bubble experiment'', which has the goal of
constructing advanced forecasts of bubbles and crashes. In the
financial bubble experiment, the results are revealed only after the
predicted event has passed but the original date when we produced
the forecasts has been publicly, digitally authenticated
\cite{BFE-FCO09,BFE-FCO10}.

\section{Conclusion}

In this paper, we discussed how leverage can influence the liquidity
of the market and used the observation that a dramatic decrease of
leverage coincides with the recent financial crisis. The market size
of repos is a very good proxy for the overall leverage of the
market. We used the JLS model of log-periodic power law dynamics on
an ensemble of intervals from a time series of the total repos
market size and found that the range of crash times $t_c$ as
forecast by the fits is consistent with the observed peak and
subsequent crash.


\clearpage
\begin{thebibliography}{19}
\expandafter\ifx\csname
natexlab\endcsname\relax\def\natexlab#1{#1}\fi
\providecommand{\bibinfo}[2]{#2} \ifx\xfnm\relax
\def\xfnm[#1]{\unskip,\space#1}\fi
\bibitem[{Dudley(2010)}]{dudley20100409}
\bibinfo{author}{W.~C. Dudley},
\newblock \bibinfo{title}{The friday podcast: New york fed chief, bubble
  fighter},
\newblock \bibinfo{journal}{NPR Planet Money}  (\bibinfo{year}{2010}).
\bibitem[{Case and Shiller(2003)}]{CaseShiller}
\bibinfo{author}{K.~E. Case}, \bibinfo{author}{R.~J. Shiller},
\newblock \bibinfo{title}{{Is there a bubble in the housing market}},
\newblock \bibinfo{journal}{Brookings Papers Econ. Activity}
  \bibinfo{volume}{(2)} (\bibinfo{year}{2003}) \bibinfo{pages}{299--362}.
\bibitem[{Johansen et~al.(1999)Johansen, Sornette, and Ledoit}]{jsl}
\bibinfo{author}{A.~Johansen}, \bibinfo{author}{D.~Sornette},
  \bibinfo{author}{O.~Ledoit},
\newblock \bibinfo{title}{Predicting financial crashes using discrete scale
  invariance},
\newblock \bibinfo{journal}{Journal of Risk} \bibinfo{volume}{1}
  (\bibinfo{year}{1999}) \bibinfo{pages}{5--32}.
\bibitem[{Johansen et~al.(2000)Johansen, Ledoit, and Sornette}]{Johansen2000f}
\bibinfo{author}{A.~Johansen}, \bibinfo{author}{O.~Ledoit},
  \bibinfo{author}{D.~Sornette},
\newblock \bibinfo{title}{{Crashes as critical points}},
\newblock \bibinfo{journal}{International Journal of Theoretical and Applied
  Finance} \bibinfo{volume}{3} (\bibinfo{year}{2000})
  \bibinfo{pages}{219--255}.
\bibitem[{wikipedia repo(page)}]{repowiki}
\bibinfo{author}{wikipedia repo},
\newblock \bibinfo{journal}{:
  http://en.wikipedia.org/wiki/Repurchase\_agreement}
  (\bibinfo{year}{webpage}).
\bibitem[{Adrian and Shin(2010)}]{liquidityleverage}
\bibinfo{author}{T.~Adrian}, \bibinfo{author}{H.-S. Shin},
\newblock \bibinfo{title}{Liquidity and leverage},
\newblock \bibinfo{journal}{Journal of Financial Intermediation}
  \bibinfo{volume}{19} (\bibinfo{year}{2010}) \bibinfo{pages}{418--437}.
\bibitem[{Sornette et~al.(1996)Sornette, Johansen, and Bouchaud}]{Sornette1996}
\bibinfo{author}{D.~Sornette}, \bibinfo{author}{A.~Johansen},
  \bibinfo{author}{J.~P. Bouchaud},
\newblock \bibinfo{title}{{Stock market crashes, precursors and replicas}},
\newblock \bibinfo{journal}{J. Phys. I France} \bibinfo{volume}{6}
  (\bibinfo{year}{1996}) \bibinfo{pages}{167--175}.
\bibitem[{Sornette(2003)}]{sornettecrash}
\bibinfo{author}{D.~Sornette},
\newblock \bibinfo{title}{Why stock markets crash (critical events in complex
  financial systems)},
\newblock \bibinfo{journal}{Why Stock Markets Crash (Critical Events in Complex
  Financial Systems), Princeton University Press}  (\bibinfo{year}{2003}).
\bibitem[{Sornette et~al.(2009)Sornette, Woodard, and Zhou}]{oil}
\bibinfo{author}{D.~Sornette}, \bibinfo{author}{R.~Woodard},
  \bibinfo{author}{W.-X. Zhou},
\newblock \bibinfo{title}{The 2006-2008 oil bubble: evidence of speculation and
  prediction},
\newblock \bibinfo{journal}{Physica A} \bibinfo{volume}{388}
  (\bibinfo{year}{2009}) \bibinfo{pages}{1571--1576}.
\bibitem[{Jiang et~al.(2010)Jiang, Zhou, Sornette, Woodard, Bastiaensen, and
  Cauwels}]{ssec}
\bibinfo{author}{Z.-Q. Jiang}, \bibinfo{author}{W.-X. Zhou},
  \bibinfo{author}{D.~Sornette}, \bibinfo{author}{R.~Woodard},
  \bibinfo{author}{K.~Bastiaensen}, \bibinfo{author}{P.~Cauwels},
\newblock \bibinfo{title}{Bubble diagnosis and prediction of the 2005-2007 and
  2008-2009 chinese stock market bubbles},
\newblock \bibinfo{journal}{Journal of Economic Behavior and Organization}
  \bibinfo{volume}{74} (\bibinfo{year}{2010}) \bibinfo{pages}{149--162}.
\bibitem[{Zhou and Sornette(2008)}]{vegas}
\bibinfo{author}{W.-X. Zhou}, \bibinfo{author}{D.~Sornette},
\newblock \bibinfo{title}{Analysis of the real estate market in las vegas:
  Bubble, seasonal patterns, and prediction of the csw indexes},
\newblock \bibinfo{journal}{Physica A} \bibinfo{volume}{387}
  (\bibinfo{year}{2008}) \bibinfo{pages}{243--260}.
\bibitem[{Zhou and Sornette(2006)}]{southafrica}
\bibinfo{author}{W.-X. Zhou}, \bibinfo{author}{D.~Sornette},
\newblock \bibinfo{title}{A case study of speculative financial bubbles in the
  south african stock market 2003-2006},
\newblock \bibinfo{journal}{Physica A} \bibinfo{volume}{361}
  (\bibinfo{year}{2006}) \bibinfo{pages}{297--308}.
\bibitem[{Yan et~al.(2010)Yan, Woodard, and Sornette}]{prrebound}
\bibinfo{author}{W.~Yan}, \bibinfo{author}{R.~Woodard},
  \bibinfo{author}{D.~Sornette},
\newblock \bibinfo{title}{Diagnosis and prediction of market rebounds in
  financial markets, international journal of forecasting},
\newblock \bibinfo{journal}{New Journal of Physics} \bibinfo{volume}{submitted}
  (\bibinfo{year}{2010}). \bibinfo{note}{Http://arxiv.org/abs/1001.0265}.
\bibitem[{Johansen and Sornette(1999)}]{js}
\bibinfo{author}{A.~Johansen}, \bibinfo{author}{D.~Sornette},
\newblock \bibinfo{title}{Critical crashes},
\newblock \bibinfo{journal}{Risk} \bibinfo{volume}{12} (\bibinfo{year}{1999})
  \bibinfo{pages}{91--94}.
\bibitem[{Cvijovic and Klinowski(1995)}]{ck}
\bibinfo{author}{D.~Cvijovic}, \bibinfo{author}{J.~Klinowski},
\newblock \bibinfo{title}{Taboo search: an approach to the multiple minima
  problem},
\newblock \bibinfo{journal}{Science} \bibinfo{volume}{267}
  (\bibinfo{year}{1995}) \bibinfo{pages}{664--666}.
\bibitem[{Levenberg(1944)}]{kl}
\bibinfo{author}{K.~Levenberg},
\newblock \bibinfo{title}{A method for the solution of certain non-linear
  problems in least squares},
\newblock \bibinfo{journal}{Quarterly Journal of Applied Mathematics II}
  \bibinfo{volume}{2} (\bibinfo{year}{1944}) \bibinfo{pages}{164--168}.
\bibitem[{Marquardt(1963)}]{dm}
\bibinfo{author}{D.~W. Marquardt},
\newblock \bibinfo{title}{An algorithm for least-squares estimation of
  nonlinear parameters},
\newblock \bibinfo{journal}{Journal of the Society for Industrial and Applied
  Mathematics} \bibinfo{volume}{11} (\bibinfo{year}{1963})
  \bibinfo{pages}{431--441}.
\bibitem[{Sornette et~al.(2010{\natexlab{a}})Sornette, Woodard, Fedorovsky,
  Reimann, Woodard, and Zhou}]{BFE-FCO09}
\bibinfo{author}{D.~Sornette}, \bibinfo{author}{R.~Woodard},
  \bibinfo{author}{M.~Fedorovsky}, \bibinfo{author}{S.~Reimann},
  \bibinfo{author}{H.~Woodard}, \bibinfo{author}{W.-X. Zhou},
\newblock \bibinfo{title}{The financial bubble experiment: advanced diagnostics
  and forecasts of bubble terminations (the financial crisis observatory)},
\newblock \bibinfo{journal}{http://arxiv.org/abs/0911.0454}
  (\bibinfo{year}{2010}{\natexlab{a}}).
\bibitem[{Sornette et~al.(2010{\natexlab{b}})Sornette, Woodard, Fedorovsky,
  Reimann, Woodard, and Zhou}]{BFE-FCO10}
\bibinfo{author}{D.~Sornette}, \bibinfo{author}{R.~Woodard},
  \bibinfo{author}{M.~Fedorovsky}, \bibinfo{author}{S.~Reimann},
  \bibinfo{author}{H.~Woodard}, \bibinfo{author}{W.-X. Zhou},
\newblock \bibinfo{title}{The financial bubble experiment: Advanced diagnostics
  and forecasts of bubble terminations volume ii--master document},
\newblock \bibinfo{journal}{http://arxiv.org/abs/1005.5675}
  (\bibinfo{year}{2010}{\natexlab{b}}).

\end{thebibliography}
\end{document}